\documentclass[12pt,preprint]{aastex}

\slugcomment{} \shorttitle{Polar BAL Outflows} \shortauthors{Zhou
et al.}

\begin{document}

\title{Polar Outflows in Six Broad Absorption Line Quasars}
\author{Hongyan Zhou\altaffilmark{1}, Tinggui Wang\altaffilmark{1}, Huiyuan Wang\altaffilmark{1},
Junxian Wang\altaffilmark{1}, Weimin Yuan\altaffilmark{2}, and Yu Lu\altaffilmark{1}}

\altaffiltext{1}{Center for Astrophysics, University of Science
and Technology of China, Hefei, Anhui, P.R.China}

\altaffiltext{2}{Yunnan Astronomical Observatory, National
Astronomical Observatories, Chinese Academy of Sciences, Kunming,
Yunnan, P.R.China}

\email{mtzhou@ustc.edu.cn}

\begin{abstract}
Using the radio observations by FIRST and NVSS, we build a sample
of 151 radio variable quasars selected from the Sloan Digital Sky
Survey Data Release 3 (SDSS DR3). Six (probably another two) among
them are classified as broad absorption line (BAL) quasars, with
radio flux variations of a few 10 percent within 1.5-5 years. Such
large amplitudes of the variations imply brightness temperatures
much higher than the inverse Compton limits (10$^{12}$ K) in all
the BAL quasars, suggesting the presence of relativistic jets
beaming toward the observer. The angle between the outflow and the
jet is constrained to be less than $\sim 20^{\circ}$. Such BAL
quasars with polar outflows are beyond the simple unification
models of BAL quasars and non-BAL quasars, which hypothesize that
BAL quasars are normal quasars seen nearly edge-on.

\end{abstract}


\section{Introduction}
The broad absorption lines (BALs), which appear in the UV spectra
of $\sim 15\%$ quasars (e.g., Tolea et al. 2002; Reichard et al.
2003a; Hewett \& Foltz 2003), are characterized by prominent broad
and blueshifted absorption troughs due to ions of a wide range
species from FeII, MgII to OVI. It is now commonly accepted that
BAL region is present in all quasars but with covering factor much
less than unit. The dichotomy of BAL quasars and non-BAL quasars
is interpreted as an orientation effect. For instance, Murray et
al. (1995) suggested that BALs present themselves in a minority of
the quasar population when the line of sight passes through the
accretion disk wind nearly at the equatorial plane. A detail study
shows that such a scenario is also consistent with the continuum
polarization and X-ray absorption in BAL quasars (Wang, Wang \&
Wang 2005). Elvis (2000) appealed to a funnel-shaped thin shell
outflow that arises from the accretion disk to explain various
observational properties of quasars, including BALs which are
interpreted as ``normal'' quasars when the wind is viewed end-on.
Both orientation models require a rather large incline angle of
$i\sim 60^{\circ}$ for BAL quasars.

The determination of the inclination of the accretion disk in BAL
quasars is vital to understand the geometry of the BAL outflow.
The axis defined by relativistic radio jets, which is likely
aligned with that of the accretion disk (Wills et al. 1999), can
be used to infer the inclination angle of the accretion disk.
Becker et al. (2000) found that the 29 BAL quasars discovered in
the FIRST Bright Quasar Survey (FBQS) exhibit compact radio
morphologies ($\sim 80\% $ unresolved at $0^{''}.2$ resolution)
and show wide scatter in the radio spectral index ($\alpha \sim
-0.7~-~1.2$ and $\alpha<0.5$ for $\sim 1/3$ sources,
$S_{\nu}\propto \nu^{-\alpha}$). According to the unification
model of radio-loud AGNs (cf. Urry \& Padovani 1995),
core-dominated flat-spectrum radio sources are those viewed close
to the radio jet axis, while lobe-dominated steep-spectrum radio
sources appear to be present at larger viewing angles. The radio
morphology and spectra of FBQS BAL quasars indicate that the
orientation of BAL quasars can be both face-on and edge-on,
contrary to the simple unification model introduced above.
However, their radio spectral indexes, derived from
non-simultaneous observations, might be biased by radio
variations. In addition, most FIRST detected BAL QSOs are only
radio-intermediate, for which it is unknown if the unification
model based on radio-loud AGN (i.e., core-dominated flat-spectrum
radio sources are face-on) still applies. Also the size of
radio-intermediate sources might be much smaller, such as the
radio intermediate quasar III ZW 2 (Brunthaler et al. 2005), thus
observations with much higher spatial resolution is required to
confirm its compactness. Jiang \& Wang (2003) acquired high
resolution VLBI images of three BAL quasars at 1.6~GHz. They found
one source is resolved into asymmetric two-sided structure and a
bright central component at $\sim 20$ mas resolution. This
morphology mimics a Compact Steep Spectral (CSS) source, but its
size is much smaller than the typical value for a CSS source. The
other two sources remain unresolved at this resolution, indicating
that they are viewed face-on.

The radio flux variability, which is sometimes a better indicator
of jet orientation than the radio morphology (at a limited
resolution and sensitivity) and the spectral slope (often based on
non-simultaneous observations), has not hitherto been adequately
explored. Becker et al. (2000) commented that 5 BAL quasars in
their sample appear to be variable at 1.4 GHz, but no further
concern was given. In this paper we use two-epoch observations at
the same frequency of 1.4 GHz by FIRST (the Faint Images of the
Radio Sky at Twenty centimeters survey, Becker et al. 1995) and
NVSS (the NRAO VLA Sky Survey, Condon et al. 1998) to study the
radio variability of the BAL quasars observed by SDSS (the Sloan
Digital Sky Survey, York et al. 2000). We identify 6 (probably
another two) BAL quasars with radio variation with a few 10
percent. Calculations based on the radio variations imply that
these sources should be viewed face-on with inclination angles
less than 20$^{\circ}$. This confirms the existence of polar BAL
outflows, opposite to the unification model for BAL quasars.
Throughout the paper, we assume a cosmology with $H_{0}$= 70 km\,
s$^{-1}$\,Mpc$^{-1}$, $\Omega_{M}=0.3$, and
$\Omega_{\Lambda}=0.7$.

\section{Data and Analysis}

\subsection{The SDSS Quasar Catalog and the Radio Data}
Our jumping-off point is the SDSS Quasar Catalog (the 3rd edition,
Schneider et al. 2005, hereafter S05C for the catalog). Its sky
coverage is $\simeq 4,188~deg^{2}$, which accounts for more than
2/5 of $\sim 10^4~deg^{2}$ planed by SDSS. S05C consists of the
46,420 quasars with $i$\-band absolute magnitudes $M_{i} < -22.0$
and positional uncertainties $< 0^{''}.2$. In particular,
optically unresolved objects brighter than $i=19^m.1$ with FIRST
counterpart within $2^{''}$ are observed, which we are more
interested in.

Using the NARO Very Large Array in its B configuration, the FIRST
survey began in 1993 was designed to explore the faint radio sky
down to a limit of $S_{1.4GHz}\sim 1 $ mJy with a resolution of
$\sim 5^{''}$ FWHM. Its sky coverage is mostly superposed with
that of SDSS. The FIRST radio catalog was presented in White et
al. (1997) with a positional accuracy of $< 1\arcsec$ (with 90\%
confidence level). By matching the SC05 and the FIRST catalog, we
find 3,757 quasars in S05C with FIRST counterparts within
$2^{''}$. Considering the average source density of $\sim 90~
deg^{-2}$ in the FIRST survey, we expect only $\sim 0.1\%$ of
these matches to be spurious. We note that the above cutoff is
biased against quasars with extended radio morphologies, e.g.,
lobe-dominated quasars. However, missing this kind of radio
variable sources based on the data presently available (see \S2.2)
and this does not influence our main results. After all, only
$\sim 8\%$ lobe-dominated radio sources will be lost (Ivezic et
al. 2002; Lu et al. in preparation), and such sources only
accounts for $\sim 3.1\%$ in a well-defined variable sample of de
Vries et al. (2004).

NVSS, using the VLA in its more compact D and DnC
configurations, was carried out between $1993-1996$ at the same
frequency as FIRST. NVSS covers all the sky area of FIRST,
but with a higher survey limit of $S_{1.4GHz}\sim 2.5$ mJy and
a lower resolution of $FWHM\sim 45^{''}$. The positional uncertainties
are estimated to be varying from $\lesssim 1^{''}$ for bright
sources with flux density $>$ 15 mJy to $7^{''}$ at the
survey limit. With a typical background noise of 0.45 mJy (about 3
times higher than that of FIRST), we expect that the NVSS survey
be able to detect all of the FIRST sources with flux
density $S_{1.4GHz}\gtrsim 5$ mJy, provided that the radio
sources do not variate.

\subsection{The Selection of Radio Variable Quasars}

We first select quasars in S05C with redshift $>$ 0.5 so that the
MgII absorption trough (or other bluer troughs), if presents,
falls within the wavelength coverage of the SDSS spectrograph
($3,800-9,200~\AA$). Out of these quasars, 1,877 have FIRST
counterparts within $2^{''}$ and peak flux density $S_{FP}
>$ 5 mJy as measured by FIRST. Then the FIRST images of all these 1,877
objects are visually inspected and classified into three
categories: 1) compact radio sources, 2) marginally resolved radio
sources, and 3) radio sources with complex morphology. The first
category includes 1,482 radio point sources unresolved at the
FIRST resolution (79.0\%). The second category includes 200 radio
sources (10.7\%), which often show core plus elongated (jet-like)
structure, or a core embedded in weak diffuse emission. The third
category includes 193 radio sources (10.7\%), out of which 168
sources exhibit Fanaroff-Riley II morphology (FR-II, Fanaroff \&
Riley 1974).

We search for NVSS counterparts for the quasars in all of the
three categories within a $21''$ matching radius ($3~\sigma$ of
the NVSS positional error at the survey limit). NVSS counterparts
are found for 1,838 of the 1,877 quasars within a $21''$ with a
false rate of less than 1\%. Two possibilities may lead unfindable
of the 39 quasars as NVSS counterparts within such a matching
radius: 1) the flux falls below the NVSS limit resulting from
variability; 2) the apparent centroid of the source is largely
shifted due to contamination either by bright lobe(s) or by nearby
unrelated bright source(s). We compare the NVSS and FIRST images
to distinguish between the two and find that all of the 39 cases
are due to confusion effects. In fact, 21 of the 39 quasars show
FR-II morphology. These 39 quasars are excluded in our further
analysis. This result indicates that care must be taken to compare
the fluxes obtained during the two surveys.

As noted by de Vries et al. (2004), the flux densities between
FIRST and NVSS are not directly comparable since the radio sources
that are not resolved by NVSS may be resolved by FIRST. In this
case, the peak flux is smaller than the integrated flux density.
We define the variability ratio (VR) for each quasar as
\begin{equation}\label{eq1}
VR=S_{FP}/S_{NI},
\end{equation}
where $S_{FP}$ and $S_{NI}$ denote the peak flux density measured
by FIRST and the integrated flux density measured by NVSS. Our
variability ratio is conservative for selecting variable sources
which are brighter in the FIRST images. We estimate the
significance of radio flux variability as
\begin{equation}\label{eq2}
\sigma_{var}=\frac{S_{FP}-S_{NI}}{\sqrt{\sigma^2_{FP}+\sigma^2_{NI}}}
\end{equation}
where $\sigma_{NI}$ is the NVSS integrated flux uncertainty, and
$\sigma_{FP}$ the FIRST peak flux uncertainty. Objects with $VR >
1$ and $\sigma_{var} > 3$ are taken as candidates radio variable
quasars. This yields 154 candidate variable sources. We plot in
Figure \ref{f1} (left panel) variability ratio VR against the peak
flux density measured by FIRST $S_{FP}$. The apparent dependence
of VR on $S_{FP}$ is due to the combination of two facts, the
dependence of measurement error on flux density and the confusion
effects in the NVSS. The latter complication also induces the
obvious asymmetric distribute of these sources around $VR = 1$. In
fact, all but three (SDSS J160953.42+433411.5, SDSS
J101754.85+470529.3, and SDSS J164928.87+304652.4) of radio
sources that are well resolved by FIRST have $VR<1$. A more
symmetric distribution can be found if only point radio sources
and marginally resolved radio sources (symbols in blue and red
color) are considered. Isolated compact sources located well below
$VR=1$ are likely radio variable quasars. But new observations
with higher resolution are needed to confirm this. Out of the
three quasars with complex radio morphology and $VR>1$ (all are
FR-II quasars), SDSS J160953.42+433411.5 has $VR = 2.26$ and
$\sigma_{var} = 20.90$, fulfilling our selection criteria. After
careful examination of its FIRST and NVSS image, we find confusion
effects in NVSS are serious in this source and exclude it from our
sample. Out of rest 153 candidates, 151 are point radio sources
and 2 are marginally resolved. Their NVSS and FIRST images are all
visually examined for possible contamination by unrelated nearby
bright sources, and one of them, SDSS J111344.84-004411.6, is
removed for this reason. In addition, SDSS J094420.44+613550.1 is
eliminated from the sample because the NVSS pipeline gives a wrong
flux density for this object. At last, we end up with a sample of
151 candidate radio variable quasars\footnote{Considering the
systematic uncertainty of radio flux observed by FIRST and NVSS
does not significantly alter the results of this paper.}. Using
the FIRST integrated flux density and the SDSS PSF magnitudes, we
calculate the radio-loudness (RL) of these quasars, defined as the
k-corrected ratio of the 5 GHz radio flux to the near ultraviolet
flux at 2 500 \AA, $RL=S_{\nu,5GHz}/S_{\nu,2500\AA}$. A power law
slope of $\alpha=0$ ($S_{\nu}\varpropto \nu^{-\alpha}$) is assumed
for radio emission since their radio spectra are likely flat (Wang
et al. 2005), and the SDSS colors are used for the optical-UV
emission. The radio properties of a sub-sample of the BAL quasars
(see \S2.3 for detailed description) selected from these objects
are listed in Table 1. In Figure \ref{f1} (right panel), we plot
the radio-loudness against radio variability ratio for the 151
quasars. We see that most radio sources with large variability
amplitude ($VR\gtrsim 1.5$) have the radio-loudness $10 \lesssim
RL \lesssim 250$.

\subsection{The Optical Spectral Analysis and the Sample of Radio Variable BAL Quasars}
\begin{figure}
\epsscale{1.0} \plotone{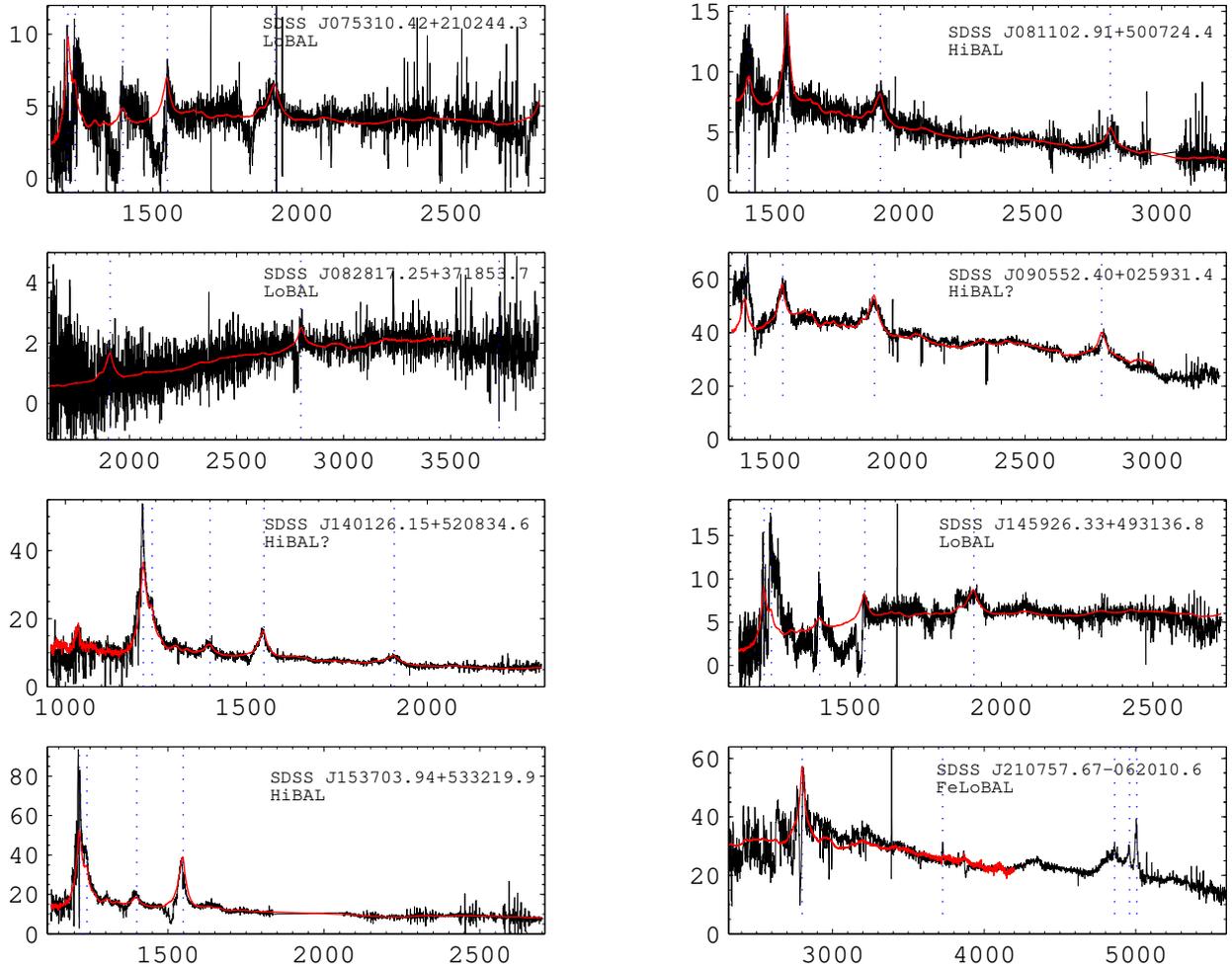} \caption{The SDSS composite quasar
spectrum fits (red curve) to the spectra of eight radio variable
BAL quasars (or candidates labelled with question masks).}
\label{f2}
\end{figure}

Eight BAL candidates were identified by visually inspecting the
SDSS spectra of the 151 radio variable quasars. These candidate
spectra were corrected for Galactic extinction using the
extinction curve of Schlegel et al. (1998) and brought to their
rest frame before further analysis. We use the ``Balnicity" Index
(BI) defined by Weymann et al. (1991) and Reichard (2003b) to
quantitatively classify the absorption troughs. Following the
procedures described by Reichard et al. (2003b), we calculated the
BIs for the broad absorption troughs by comparing the observed
spectra of our BAL candidates with the composite quasar spectrum
created by Vanden Berk et al. (2001) from the SDSS Early Data
Release (EDR). In brief, the EDR composite spectrum was reddened
using the Pei (1992) SMC extinction curve to match each observed
spectrum of our BAL candidates. The fit is done through
minimization of $\chi^2$ with proper weights given to the
wavelength regions that are obviously affected by prominent
emission lines and absorption troughs. The results are displayed
in Figure \ref{f2}. Then we calculated the balnicity index of high
ionization line BI(CIV) using the definition of Weymann et al.
(1991), and the balnicity index of low ionization line BI(MgII)
and BI(AlIII) using the definition of Reichard (2003b):
\begin{equation}\label{eq3}
BI=\int^{25,000}_{0~or~3,000}dv[1-\frac{F^{obs}(v)}{0.9F^{fit}(v)}]C(v)
\end{equation}
where $F^{obs}(v)$ and $F^{fit}(v)$ are the observed and fitted
fluxes as a function of velocity in km~s$^{-1}$ from the systemic
redshift within the range of each absorption trough, and
\begin{equation}\label{eq4}
C(v)=\left\{%
\begin{array}{ll}
    1.0, & {\rm if~[1-\frac{F^{obs}(v)}{0.9F^{fit}(v)}]>0~ over~a~continous~interval~of~\gtrsim W~km~s^{-1},}\\
    0, & {\rm otherwise.} \\
\end{array}%
\right.
\end{equation}
The integral in Equation \ref{eq3} begins at $v=3,000~km~s^{-1}$
for CIV and at $v=0~km~s^{-1}$ for MgII and AlIII. The given
continuous interval  in Equation \ref{eq4} is $W=2,000~km~s^{-1}$
for CIV and $W=1,000~km~s^{-1}$ for MgII and AlIII. The results
are listed in Table \ref{t1} and described individually below:
\begin{description}
  \item[SDSS J075310.42$+$210244.3]Apart from the deep high-ionization BAL
troughs of CIV, SiIV, NV with $v\sim$ 0-13,500~km~s$^{-1}$, the low-ionization
AlIII trough, covering a similar velocity range, is also obvious. At the red
end of the spectrum, the MgII BAL trough is apparently present. With
$BI$(CIV)=3,633~km~s$^{-1}$ and $BI$(AlIII)=1,420~km~s$^{-1}$, we are
certainly observing a LoBAL quasar in this object.
  \item[SDSS J081102.91$+$500724.4]Based on the Balnicity Index of
$BI(CIV)=617~km~s^{-1}$ and the velocity range of $v\sim 6,700-11,600~km~s^{-1}$, this object can be safely classified
  as a HiBAL quasar. Much shallower low-ionization BAL troughs of AlIII and MgII with
  similar velocity range may also be present. However, high S/N
  spectrum is needed to confirm this.
  \item[SDSS J082817.25$+$371853.7]Though its SDSS spectrum is rather noisy, the low-ionization BAL
  troughs of AlIII and MgII are securely identified. Classification of
  this object as LoBAL quasar should be safe based on our conservatively calculated
  BI of $1,890~km~s^{-1}$ for AlIII and $626~km~s^{-1}$ for MgII.
  \item[SDSS J090552.40$+$025931.4]A BAL trough detached
$\sim 20,000~km~s^{-1}$ from the CIV peak is recognizable. However, the fit
with the composite quasar spectrum yields $BI(CIV)=0~km~s^{-1}$. A marginal
value of BI(CIV)=228~km~s$^{-1}$ is obtained if using a power law to fit the
continuum. We tentatively classified this object as a candidate HiBAL quasar.
  \item[SDSS J140126.15$+$520834.6]This object shows rather shallow BAL
trough bluewards of CIV emission line. We label it as a candidate.
  \item[SDSS J145926.33$+$493136.8]This object show two sets of BAL troughs,
SiIV, CIV, and AlIII, around velocities $v\sim$ 4,000~km~s$^{-1}$
and $v\sim$ 15,000~km~s$^{-1}$. We classified it as a LoBAL solely
by the presence of AlIII BAL troughs, since MgII is redshifted out
of the wavelength coverage of the SDSS spectrograph. Broad
$Ly\alpha$ emission line is completely absorbed and only narrow
$Ly\alpha$ emission line appears in the spectrum. Broad NV
emission line is extremely strong perhaps due to scattering the
$Ly\alpha$ emission line (Wang, Wang, \& Wang, in preparation).
  \item[SDSS J153703.94$+$533219.9]We classified this object as a
HiBAL quasar based on the presence of detached CIV BAL trough with
$BI(CIV)=2,060~km~s^{-1}$. AlIII falls in the wavelength range of
bad pixel and MgII is redshifted out of spectroscopic coverage.
  \item[SDSS J210757.67$-$062010.6]This object belong to the rare
class of FeLoBAL quasars (e.g., Becker et al. 1997) characterized
by the metastable FeII BAL troughs centered at 2,575 \AA~ and MgII
BAL troughs. Many FeII absorption features are also detected
redward of MgII, as well as neutral helium absorption triplet,
$HeI\lambda\lambda2946,3189,3890$, which can be used as powerful
diagnostics of HI column density (Arav et al. 2001 and references
therein). Only a few quasars have been found to show HeI
absorption lines (e.g., Anderson 1974; Arav et al. 2001; Hall et
al. 2002).
\end{description}

\section{Discussion and Conclusion}

\begin{figure}
\epsscale{1.0} \plotone{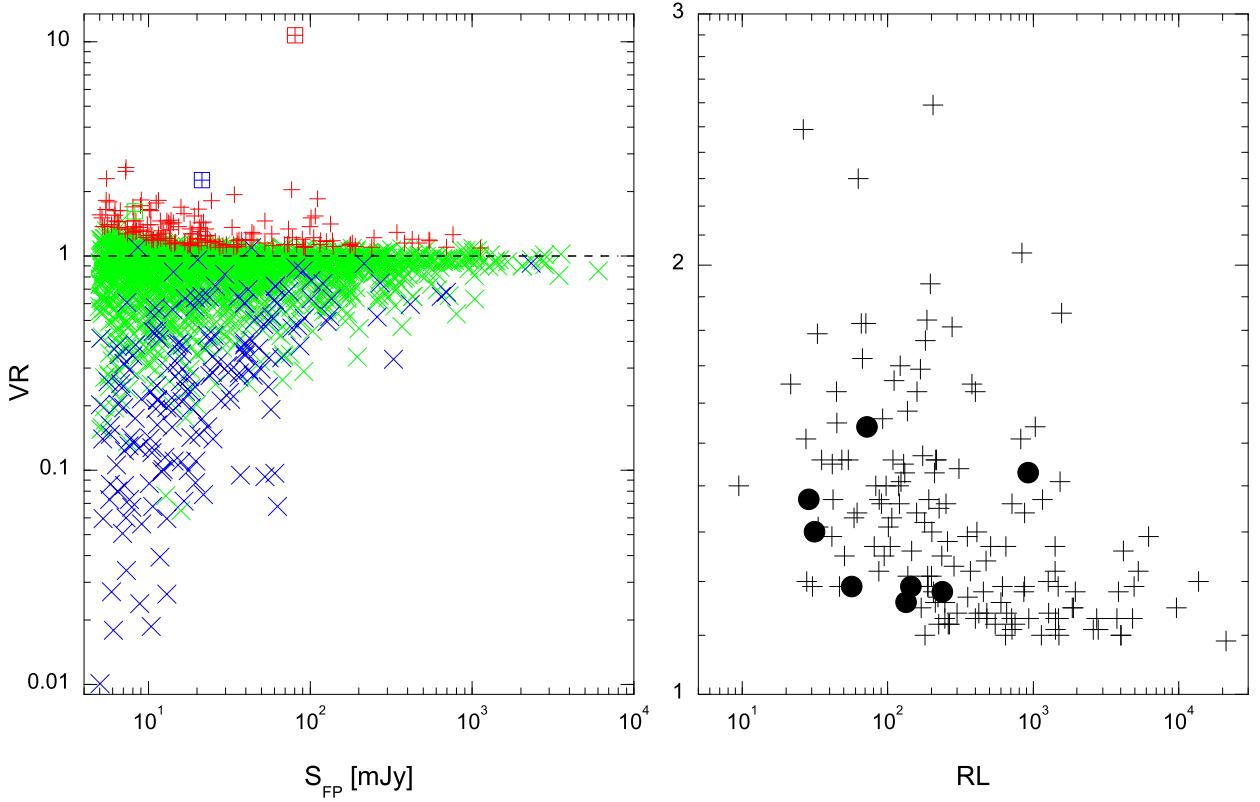} \caption{The left panel shows the
distribution of the 1,838 radio quasars on the flux density
variation VR versus flux density $S_{FP}$ plan. VR is defined as
as the ratio of peak flux density measured by FIRST ($S_{FP}$) to
integrated flux density measured by NVSS. Objects in the final
radio variable sample are denoted as red ``+". Objects with
complex morphology are denoted as blue ``$\times $". Objects that
are unresolved or marginally resolved by the FIRST are denoted as
green ``$\times $". Objects that fulfill the definition criteria
but have been excluded from our sample due to contamination by
nearby component(s) or unrelated sources are denoted as ``+" with
a box envelope in corresponding color box (see the text for
detailed description). Note that, due to confusion effects in the
NVSS, the asymmetric distribution of these quasars around $VR=1$
and this asymmetry is more serious for lobe-dominated sources than
unresolved and marginally resolved sources. The flux density
variation VR is plotted against radio-loudness RL for our sample
of 151 radio variable quasars in the right panel. Normal quasars
are denoted ``+" and BAL quasars by solid circles.} \label{f1}
\end{figure}

The origin of radio variability can be either extrinsic or
intrinsic. The most familiar mechanism of extrinsic variability of
radio sources is refractive InterStellar Scintillation (ISS, e.g.,
Blandford, Narayan, \& Romani 1986). However, the variability amplitudes
induced by ISS is seldom larger than $\sim 2\%$. Considering the
fact that all the 6 BAL quasars and the two BAL candidates we selected
show radio variation with amplitude $>$ $15\%$
(Figure \ref{f1} and Table \ref{t1}), it is reasonable to believe
that the radio variabilities of these BAL quasars are intrinsic to the
radio sources. Marscher \& Gear (1985) suggested that shocks
propagating along the radio jets can induce flux variability.
Amplitude of variability can be largely amplified by relativistic
beaming effect if the radio jets are viewed nearly face-on.

A lower limit of the brightness temperature can be inferred as
follows (Krolik 1999),
\begin{equation}\label{eq5}
T_{B}^{l}\sim \frac{\Delta P_{\nu}}{2k_{B}\nu^2\Delta t^2},
\end{equation}
where $\Delta P_{\nu}$ is the variable part of the radio power
computed from the difference between the FIRST and NVSS fluxes,
$\Delta t$ the time interval in the source rest frame between two
observations, and $k_B$ the Boltzmann constant. We present
$T_{B}^{l}$ for the 6 BAL quasars and the 2 candidates in Table
\ref{t1}. We find that the brightness temperatures of all 8 radio
sources are much larger than the inverse Compton limit of
$10^{12}$ K (Kellermann \& Pauliny-Toth 1969). Such extremely
large brightness temperatures strongly suggest the presence of
relativistic jet beaming toward the observer. If the intrinsic
brightness temperature of the radio sources is less than the
inverse Compton limit, we can set a lower limit on their Doppler
factor, $\delta_{l}=(\frac{T_{B}^{l}}{10^{12~Kelvin}})^{1/3}$, and
hence an upper limit of the inclination angle,
\begin{equation}\label{eq6}
\theta_{l}=arccos\{[1-(\gamma\delta_{l})^{-1}]\beta^{-1}\},
\end{equation}
where $\gamma = (1-\beta^2)^{-1/2}$ is the Lorentz factor of the
jets. We find that all these radio variable BAL quasars/candidates must be
viewed within $\theta \lesssim 10^{\circ}$, except SDSS
J210757.67$-$062010.6, for which $\theta \lesssim 20^{\circ}$. If
the equipartition value of $\sim 5\times 10^{10}$ K (e.g.,
Readhead 1994; L{\" a}hteenm{\" a}ki et al. 1999) instead of the
inverse Compton value of $10^{12}$ K is adopted as the maximum
intrinsic brightness temperature, the inclination angle of our
sample of BAL quasars should all be less than $\sim 7^{\circ}$.
Therefore polar BAL outflows must be present in these radio
variable quasars, contrary to the simple unification models of BAL
and non-BAL quasars, which hypothesize BAL quasars are normal
quasars seen nearly edge-on.

Similar to most FIRST detected SDSS BAL quasars (Menou et al.
2001), All but one of our radio variable BAL quasars are all
radio-intermediate with $RL \lesssim 250$. With $RL=923$, SDSS
J082817.25+371853.7 is the only exceptionally very radio-loud BAL
quasar in our sample. We note, however, its spectrum is
significantly reddened with $E(B-V)\simeq 1^m$. Correcting for
this intrinsic extinction would place the quasar in the
radio-intermediate range with $RL\simeq 10$. Since the radio
emission is likely boosted greatly by the relativistic motion of
the jet as indicated by its large brightness temperature, most of
these sources would be intrinsically radio-weak or at most
radio-intermediate. The properties of jets in radio-intermediate
quasars are not well understood, but at least in one of such
object III ZW 2, super-luminal motion of knots at parsec scale has
been detected (Brunthaler et al. 2005). Recurrent radio flares was
observed in the same object. Miller et al (1993) speculated that
radio-intermediate quasars are actually relativistically boosted
radio-quiet quasars based on the correlation between [OIII]
luminosity and radio power. It is possible that the BAL gas in
these objects associated with the expansion of radio plasma. The
origin of BAL in these objects may be different from the majority
of BAL QSOs, in which a disk-wind is responsible for the
absorption.

\begin{figure}
\epsscale{1.0} \plotone{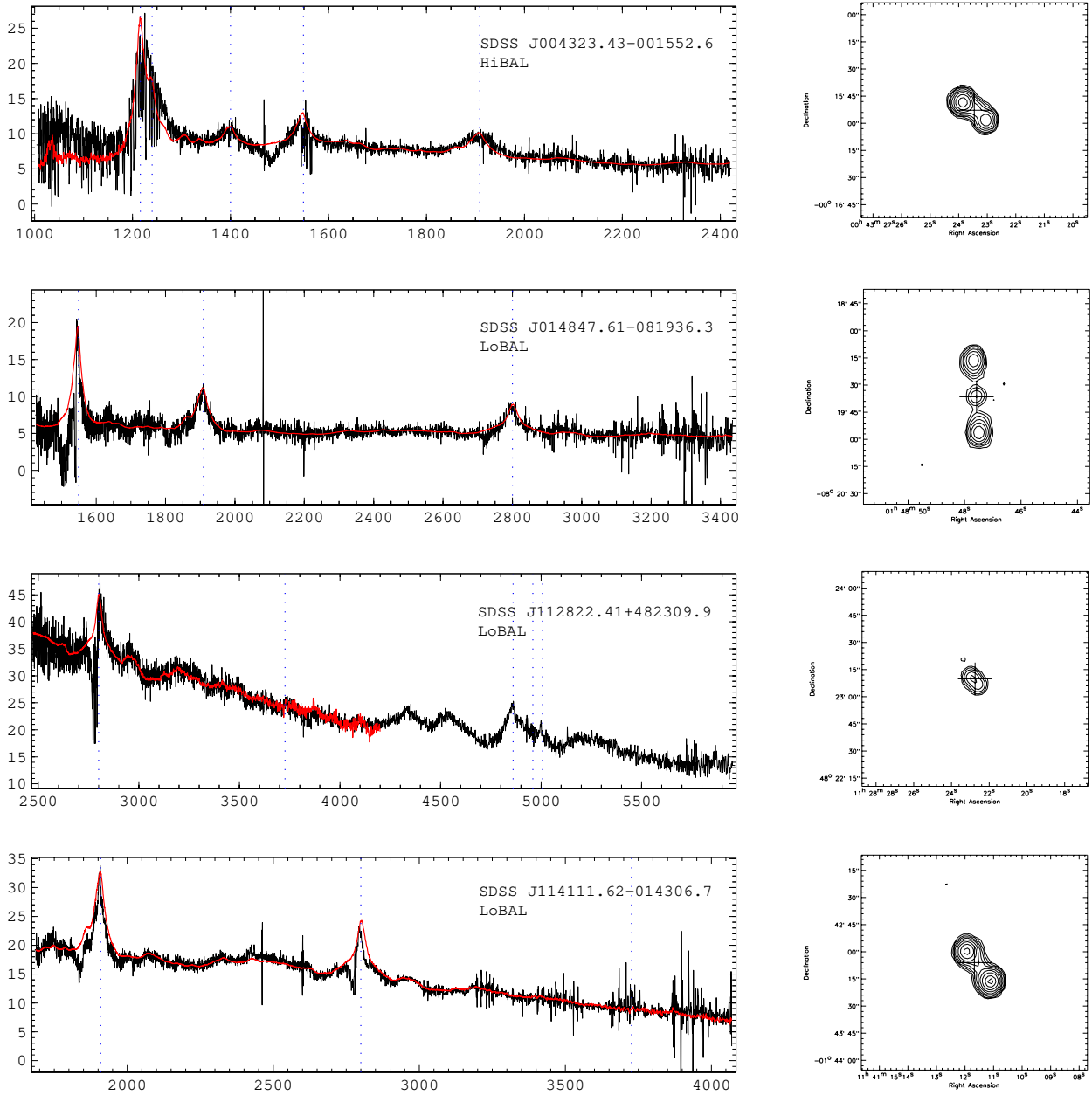} \caption{Left panels: the SDSS
composite quasar spectrum fits (red curve) for four BAL quasar
spectra, whose radio emission is resolved by FIRST. Right panels:
FIRST survey 1.4 GHz images of the corresponding BAL quasars;
contours are spaced by factors of 2 from 0.5 mJy.} \label{f3}
\end{figure}

Reichard et al. (2003a) found an uncorrected fraction of
14.0$\pm$1.0\% of BAL quasars in SDSS EDR within the redshift
ranges $1.7\le z\le 3.4$, whereas Menou et al. (2001) found this
fraction falls to 3.3$\pm$1.1\% amongst those quasars with FIRST
detection. Among the 82 radio variable QSOs within the same
redshift range in our sample, we identify 4-6 BAL quasars. The
overall fraction ($\sim$ 4\%) of BAL quasars in our radio variable
sample is similar to the fraction of the SDSS quasars detected by
FIRST, but much lower (significant at 1-2$\sigma$ level) than the
fraction of the whole SDSS quasars, the majority of which are
selected according to optical color. As a comparison, out of $\sim
600$ quasars with $z>0.5$ in the SDSS DR3 that are resolved by
FIRST image, we only identified four BAL quasars. Their SDSS
spectra with our model fits are displayed in Figure \ref{f3}, and
their FIRST images as contour maps. Three show FR-II morphology,
out of which one object, SDSS J114111.62-014306.7 (also known as
LBQS 1138-0126), was previously known as such (Brotherton et al.
2002). Apart from LBQS 1138-0126, the only other known FR-II BAL
quasar is FIRST J101614.3+520916, which was also discovered from
the FIRST survey (Gregg et al. 2000). At the sensitivity of the
FIRST image, the majority of the resolved high redshift quasars
are classical radio loud sources. The occurrence of BALs in such
radio powerful quasars is extremely small (0.7\%).

\acknowledgments We thank the anonymous referee for useful
suggestions. This work was supported by Chinese NSF through
NSF-10233030 and NSF-10473013, the Bairen Project of CAS, and a
key program of Chinese Science and Technology Ministry. This paper
has made use of the data from the SDSS. Funding for the creation
and the distribution of the SDSS Archive has been provided by the
Alfred P. Sloan Foundation, the Participating Institutions, the
National Aeronautics and Space Administration, the National
Science Foundation, the U.S. Department of Energy, the Japanese
Monbukagakusho, and the Max Planck Society. The SDSS is managed by
the Astrophysical Research Consortium (ARC) for the Participating
Institutions. The Participating Institutions are The University of
Chicago, Fermilab, the Institute for Advanced Study, the Japan
Participation Group, The Johns Hopkins University, Los Alamos
National Laboratory, the Max-Planck-Institute for Astronomy
(MPIA), the Max-Planck-Institute for Astrophysics (MPA), New
Mexico State University, Princeton University, the United States
Naval Observatory, and the University of Washington.

\begin{deluxetable}{rrrrlrrlrrrrrrl}
\setlength{\tabcolsep}{0.02in} \tablecaption{Properties of the Radio
Variable BAL Quasars\label{t1}} \rotate
\tabletypesize{\scriptsize}
\tablewidth{550.00000pt}
\tablehead{ \colhead{Coordinates (J2000)} & \colhead{Redshift} &
\colhead{$S_{F}^{p}$} & \colhead{$S_{F}^{i}$} & \colhead{Date-Obs
(F) } & \colhead{$S_{N}^{p}$} & \colhead{$S_{N}^{i}$} &
\colhead{Date-Obs (N)} & \colhead{Sig} & \colhead{$T_{b}$} &
\colhead{R} & \colhead{} & \colhead{$BI~(km~s^{-1})$} &
\colhead{} & \colhead{BAL Type}\\
\colhead{hhmmss.ss$\pm$ddmmss.s} & \colhead{} & \colhead{mJy} &
\colhead{mJy} & \colhead{yy-mm-dd} & \colhead{mJy} & \colhead{mJy}
& \colhead{yy-mm-dd} & \colhead{$\sigma$} & \colhead{Kelvin} &
\colhead{} & \colhead{CIV} & \colhead{AlIII} & \colhead{MgII} &
\colhead{}\\
\colhead{(1)} & \colhead{(2)} & \colhead{(3)} & \colhead{(4)} &
\colhead{(5)} & \colhead{(6)} & \colhead{(7)} & \colhead{(8)} &
\colhead{(9)} & \colhead{(10)} & \colhead{(11)} & \colhead{(12)} &
\colhead{(13)} & \colhead{(14)} & \colhead{(15)} }

\startdata

075310.42+210244.3&2.2918&16.78&17.52&1998-09&14.08&14.40&1993-11-01&3.64&$ 10^{13.9}$ &133&3633&1420&$-1$\tablenotemark{a}&LoBAL \\
081102.91+500724.5&1.8376&23.07&24.93&1997-05-23&18.82&19.50&1993-11-15&4.59&$ 10^{14.1}$ &237&617&0&25&HiBAL \\
082817.25+371853.7&1.3530&21.18&21.42&1994-07-23&14.53&14.80&1993-12-15&9.67&$ 10^{15.6}$ &923&$-1$\tablenotemark{a}&1890&626&LoBAL \\
090552.40+025931.4&1.8183&43.54&43.84&1998-07&35.57&36.40&1993-11-15&5.53&$ 10^{14.1}$ &56&0\tablenotemark{b}&0&0&HiBAL? \\
140126.15+520834.6&2.9724&36.18&37.13&1997-05-04&29.94&30.40&1993-11-15&5.35&$ 10^{14.8}$ &145&85&0&$-1$\tablenotemark{a}&HiBAL? \\
145926.33+493136.8&2.3700&5.22&4.74&1997-04-17&3.61&3.80&1995-03-12&3.16&$ 10^{14.4}$ &29&9039&329&$-1$\tablenotemark{a}&LoBAL \\
153703.94+533219.9&2.4035&9.28&9.58&1997-05&6.91&7.10&1993-11-15&4.81&$ 10^{14.2}$ &31&2060&$-1$\tablenotemark{a}&$-1$\tablenotemark{a}&HiBAL \\
210757.67$-$062010.6&0.6456&19.21&19.61&1997-02&12.07&12.40&1993-09-20&10.37&$10^{13.3}$
&72&$-1$\tablenotemark{a}&$-1$\tablenotemark{a}&$\gtrsim
1431$&FeLoBAL

\enddata
\tablenotetext{a}{Absorption trough falls out of the wavelength
coverage of the SDSS spectrograph.}

\tablenotetext{b}{$BI=228~km~s^{-1}$ if calculated using a power
law fit to the continuum.}

\tablecomments{Col.3: FIRST Peak flux density, Col.4: FIRST
Integrated flux density, Col.5: FIRST observation time, Col.6:
NVSS Peak flux density, Col.7: NVSS Integrated flux density,
Col.8: NVSS observation time, Col. 9: Significance of radio
variability as defined in \S2.2, Col. 10: lower limits of
brightness temperature estimated as in \S3, Col. 11:
Radio-Loudness calculated in \S2.2, Col. 12 to Col. 14: Balnicity
Index values of CIV, AlIII and MgII, respectively, Col 15: our
final classifications, question marked ones denote candidates.}
\end{deluxetable}
\clearpage

\begin{thebibliography}{}

\bibitem[Abazajian et al.(2005)]{2005AJ....129.1755A} Abazajian, K., et
al.\ 2005, \aj, 129, 1755

\bibitem[Anderson(1974)]{1974ApJ...189..195A} Anderson, K.~S.\ 1974, \apj,
189, 195

\bibitem[Arav et al.(2001)]{2001ApJ...546..140A} Arav, N., Brotherton,
M.~S., Becker, R.~H., Gregg, M.~D., White, R.~L., Price, T., \&
Hack, W.\ 2001, \apj, 546, 140


\bibitem[Becker et al.(1995)]{1995ApJ...450..559B} Becker, R.~H., White,
R.~L., \& Helfand, D.~J.\ 1995, \apj, 450, 559

\bibitem[Becker et al.(1997)]{1997ApJ...479L..93B} Becker, R.~H., Gregg,
M.~D., Hook, I.~M., McMahon, R.~G., White, R.~L., \& Helfand,
D.~J.\ 1997, \apjl, 479, L93

\bibitem[Becker et al.(2000)]{2000ApJ...538...72B} Becker, R.~H., White,
R.~L., Gregg, M.~D., Brotherton, M.~S., Laurent-Muehleisen, S.~A.,
\& Arav, N.\ 2000, \apj, 538, 72

\bibitem[Blandford et al.(1986)]{1986ApJ...301L..53B} Blandford, R.,
Narayan, R., \& Romani, R.~W.\ 1986, \apjl, 301, L53

\bibitem[Briggs et al.(1984)]{1984ApJ...287..549B} Briggs, F.~H., Turnshek,
D.~A., \& Wolfe, A.~M.\ 1984, \apj, 287, 549

\bibitem[Brotherton et al.(2002)]{2002AJ....124.2575B} Brotherton, M.~S.,
Croom, S.~M., De Breuck, C., Becker, R.~H., \& Gregg, M.~D.\ 2002, \aj,
124, 2575

\bibitem[Condon et al.(1998)]{1998AJ....115.1693C} Condon, J.~J., Cotton,
W.~D., Greisen, E.~W., Yin, Q.~F., Perley, R.~A., Taylor, G.~B.,
\& Broderick, J.~J.\ 1998, \aj, 115, 1693

\bibitem[de Vries et al.(2004)]{2004AJ....127.2565D} de Vries, W.~H.,
Becker, R.~H., White, R.~L., \& Helfand, D.~J.\ 2004, \aj, 127,
2565

\bibitem[Elvis(2000)]{2000ApJ...545...63E} Elvis, M.\ 2000, \apj, 545, 63

\bibitem[Fanaroff \& Riley(1974)]{1974MNRAS.167P..31F} Fanaroff, B.~L., \&
Riley, J.~M.\ 1974, \mnras, 167, 31P

\bibitem[Gregg et al.(2000)]{2000ApJ...544..142G} Gregg, M.~D., Becker,
R.~H., Brotherton, M.~S., Laurent-Muehleisen, S.~A., Lacy, M., \&
White, R.~L.\ 2000, \apj, 544, 142

\bibitem[Hall et al.(2002)]{2002ApJS..141..267H} Hall, P.~B., et al.\ 2002,
\apjs, 141, 267

\bibitem[Hewett \& Foltz(2003)]{2003AJ....125.1784H} Hewett, P.~C., \&
Foltz, C.~B.\ 2003, \aj, 125, 1784

\bibitem[Ivezi{\'c} et al.(2002)]{2002AJ....124.2364I} Ivezi{\'c}, {\v Z}.,
et al.\ 2002, \aj, 124, 2364

\bibitem[Kellermann \& Pauliny-Toth(1969)]{1969ApJ...155L..71K} Kellermann,
K.~I., \& Pauliny-Toth, I.~I.~K.\ 1969, \apjl, 155, L71

\bibitem[Krolik(1999)]{1999agnc.book.....K} Krolik, J.~H.\ 1999, Active
galactic nuclei : from the central black hole to the galactic
environment /Julian H.~Krolik.~Princeton, N.~J.~: Princeton
University Press

\bibitem[Jiang \& Wang(2003)]{2003A&A...397L..13J} Jiang, D.~R., \& Wang,
T.~G.\ 2003, \aap, 397, L13

\bibitem[L{\" a}hteenm{\" a}ki et al.(1999)]{1999ApJ...511..112L} L{\"
a}hteenm{\" a}ki, A., Valtaoja, E., \& Wiik, K.\ 1999, \apj, 511,
112

\bibitem[Lynds(1967)]{1967ApJ...147..396L} Lynds, C.~R.\ 1967, \apj, 147,
396

\bibitem[Marscher \& Gear(1985)]{1985ApJ...298..114M} Marscher, A.~P., \&
Gear, W.~K.\ 1985, \apj, 298, 114

\bibitem[Menou et al.(2001)]{2001ApJ...561..645M} Menou, K., et al.\ 2001,
\apj, 561, 645

\bibitem[Murray et al.(1995)]{1995ApJ...451..498M} Murray, N., Chiang, J.,
Grossman, S.~A., \& Voit, G.~M.\ 1995, \apj, 451, 498

\bibitem[Pei(1992)]{1992ApJ...395..130P} Pei, Y.~C.\ 1992, \apj, 395, 130

\bibitem[Readhead(1994)]{1994ApJ...426...51R} Readhead, A.~C.~S.\ 1994,
\apj, 426, 51

\bibitem[Reichard et al.(2003)]{2003AJ....126.2594R} Reichard, T.~A., et
al.\ 2003a, \aj, 126, 2594

\bibitem[Reichard et al.(2003)]{2003AJ....125.1711R} Reichard, T.~A., et
al.\ 2003b, \aj, 125, 1711

\bibitem[Schlegel et al.(1998)]{1998ApJ...500..525S} Schlegel, D.~J.,
Finkbeiner, D.~P., \& Davis, M.\ 1998, \apj, 500, 525

\bibitem[Schneider et al.(2005)]{511} Schneider, D.~P.\ 2005, \aj, in
press; astro-ph/0503679; S05

\bibitem[Tolea et al.(2002)]{2002ApJ...578L..31T} Tolea, A., Krolik, J.~H.,
\& Tsvetanov, Z.\ 2002, \apjl, 578, L31

\bibitem[Urry \& Padovani(1995)]{1995PASP..107..803U} Urry, C.~M., \&
Padovani, P.\ 1995, \pasp, 107, 803

\bibitem[Vanden Berk et al.(2001)]{2001AJ....122..549V} Vanden Berk, D.~E.,
et al.\ 2001, \aj, 122, 549

\bibitem[Wang et al.(2005)]{523} Wang, T., Zhou, H., Wang, J., Lu, Y. \& Lu, Y.\ 2005,
ApJ submitted

\bibitem[Wang, Wang, \& Wang (2005)]{526} Wang, H., Wang, T., \& Wang, J.\ 2005,
\apj, in press

\bibitem[Weymann et al.(1991)]{1991ApJ...373...23W} Weymann, R.~J., Morris,
S.~L., Foltz, C.~B., \& Hewett, P.~C.\ 1991, \apj, 373, 23

\bibitem[White et al.(1997)]{1997ApJ...475..479W} White, R.~L., Becker,
R.~H., Helfand, D.~J., \& Gregg, M.~D.\ 1997, \apj, 475, 479

\bibitem[Wills et al.(1999)]{1999ApJ...520L..91W} Wills, B.~J., Brandt,
W.~N., \& Laor, A.\ 1999, \apjl, 520, L91

\bibitem[York et al.(2000)]{2000AJ....120.1579Y} York, D.~G., et al.\ 2000,
\aj, 120, 1579


\end{thebibliography}
\end{document}